\documentclass[aps,pra,floatfix,reprint,nofootinbib]{revtex4-1}

\usepackage{graphicx}
\usepackage{dcolumn}
\usepackage{bm}
\usepackage{amsmath}
\usepackage{rotating}
\usepackage{amsmath,amsthm,amssymb,mathtools}
\usepackage{verbatim}
\usepackage{float}
\usepackage{subfigure}
\usepackage{outline}
\usepackage{csquotes}
\usepackage{tikz}
\usetikzlibrary{shapes.geometric, arrows}
\usepackage{braket}
\usepackage{colonequals}
\usepackage{textcomp}
\usepackage[inline]{enumitem}

\newlist{inlinelist}{enumerate*}{1}
\setlist*[inlinelist,1]{%
  label=(\roman*),
}

\DeclareMathOperator{\diag}{diag}

\begin{abstract}

\end{abstract}

\begin{document}
\title{A Hybrid Classical/Quantum Approach for Large-Scale Studies of Quantum Systems with Density Matrix Embedding Theory}
\author{Nicholas C.\ Rubin}
\affiliation{Rigetti Quantum Computing, 775 Heinz Ave., Berkeley, CA, 94710}
\email{\url{nick@rigetti.com}}
\begin{abstract}
Determining ground state energies of quantum systems by hybrid classical/quantum methods has emerged as a promising candidate application for near-term quantum computational resources. Short of large-scale fault-tolerant quantum computers, small-scale devices can be leveraged with current computational techniques to identify important subspaces of relatively large Hamiltonians. Inspired by the work that described the merging of dynamical mean-field theory (DMFT) with a small-scale quantum computational resource as an impurity solver~[Bauer et al., arXiv:1510.03859v2], we describe an alternative embedding scheme, density matrix embedding theory (DMET), that naturally fits with the output from the variational quantum eigensolver and other hybrid approaches. This approach is validated using a quantum abstract machine simulator~[Smith~et al., arXiv:1608.03355] that reproduces the ground state energy of the Hubbard model converged to the infinite limit.
\end{abstract}
\maketitle
\section{Introduction}
Many current computational studies of strongly correlated chemical or material systems involve a high accuracy simulation of their electronic structure.   A ubiquitous theme in classical computational methods is to identify a relevant subspace of the full Hamiltonian such that a high fidelity simulation in this subspace results in a wavefunction with a high degree of overlap with the true full system wavefunction.  Examples of this methodology are active space methods, popular in quantum chemistry, and embedding or impurity problems, prevalent in studies of strongly correlated materials.   The embedded problem commonly requires a full-configuration interaction-like solution (FCI, an exact diagonalization of the embedded Hamiltonian) to significantly improve the description of the electronic structure.  Unfortunately, FCI scales exponentially in the number of orbitals, restricting its application to small problems.

The integration of active space or impurity methods with quantum computation offers a route to improving wavefunction approximations by increasing the size of the active space to include relevant low-energy states.  Bauer et al.\ \cite{bauer2015hybrid} recently proposed the integration of dynamical mean-field theory (DMFT) and the quantum phase-estimation algorithm which was demonstrated in a proof-of-concept study by Kreula et al.~\cite{kreula2016few}.  In a similar vein, Bravyi and Gosset studied the complexity of impurity methods integrated with quantum computation~\cite{ bravyi2016complexity}.    An integration of complete active space self-consistent field (CASSCF) and the phase-estimation algorithm is an example of one such integration that has already been suggested~\cite{reiher2016elucidating}.

 The variational quantum eigensolver (VQE) is a hybrid classical/quantum algorithm to approximate the ground state eigenvalues and density matrices of a Hamiltonian~\cite{peruzzo2014variational, mcclean2016theory, mcclean2015algorithms, mcclean2016hybrid, wang2015quantum, PhysRevA.92.042303}. 
 The VQE is a classical optimization loop invoking a quantum abstract machine (QAM)~\cite{smith2016practical} used for state preparation and operator measurement.  Recent work investigating this approach yielded quantum circuits for ansatz state preparation that are short and demonstrated that the algorithm is potentially robust to standard noise models~\cite{mcclean2016hybrid, jctc.12.3097.Sawaya.AspuruGuzik}.

The VQE using a wavefunction ansatz that is exponentially expensive on a classical computer, such as unitary coupled-cluster~\cite{mcclean2016theory} or approximate Hamiltonian evolution~\cite{PhysRevA.92.042303}, offers a potentially advantageous alternative to other high fidelity solvers (FCI, density matrix renormalization group, etc.) for active space calculations~\cite{reiher2016elucidating} and embedding schemes~\cite{bauer2015hybrid, PhysRevA.93.032303}.  In general, the VQE can be integrated into any computational chemistry methodology that requires a high-fidelity solver with an interface requiring the one-particle and two-particle reduced density matrices ($1$- and $2$-RDM, respectively)~\cite{helgaker2014molecular, jcp.10.1063.Werner, jcp.10.1063Gidofalvi}.  

Density matrix embedding theory (DMET) is a powerful alternative to DMFT that replaces the problem of finding the impurity's local Green's function with determining the embedded problem's $1$-RDM~\cite{PhysRevLett.109.186404, jctc.301044e.Kinzia.Chan, jctc.6b00316.Wouters.Chan, PhysRevB.89.035140, jcp.141.054113.Ireneusz.Scuseria, bulik2015electron}.  DMET maps the problem of finding the ground state of a large $N$-particle system onto many interacting impurity problems.  This is achieved by self-consistently matching the $1$-RDM from a low-level wavefunction for the entire system with the $1$-RDMs of the impurities calculated with a high-level technique.  Though it is not necessarily clear that such an iterative scheme would result in accurate local RDMs, DMET has been demonstrated to be highly accurate for one- and two-dimensional lattice models~\cite{PhysRevLett.109.186404, PhysRevB.89.035140}, extended chemical systems~\cite{PhysRevB.93.035126, jcp.141.054113.Ireneusz.Scuseria, jctc.301044e.Kinzia.Chan}, and for ground state and transition state molecules~\cite{jctc.6b00316.Wouters.Chan}.  Furthermore, spectral functions (on the real-frequency axis) are accessible from DMET without any bath discretization error allowing for the construction of arbitrary dynamic correlation functions~\cite{booth2015spectral}.  These studies demonstrate the amazing quality of DMET with modestly sized impurities.

A key component of DMET is the embedded problem solver.  Ideally, this solver produces a high-accuracy $1$-RDM for the embedded problem such that the fragment piece of the mean-field $1$-RDM can be matched to the correlated $1$-RDM for the fragment.  In this work we propose the use of VQE for the embedded problem solver as an alternative to FCI or DMRG routines.  This change potentially allows for
\begin{inlinelist}
\item the study of a larger embedding Hamiltonian, allowing access to longer range correlation functions, and
\item generally accelerates the DMET algorithm by quickly finding the solution to the embedded problem on a quantum computer.
\end{inlinelist}
As a proof of principle, the ground state energy of the repulsive-$U$ Hubbard model is determined for small rings and the thermodynamic limit.
\section{Density Matrix Embedding}
In this section we review the salient features of density matrix embedding and discuss how the method naturally incorporates the use of the variational quantum eigensolver as an embedded Hamiltonian solver. References~\cite{jctc.6b00316.Wouters.Chan, bulik2015electron} provide a more detailed review on density matrix embedding while \cite{PhysRevB.89.035140, jcp.141.054113.Ireneusz.Scuseria} present alternatives to the self-consistent procedure that appears in DMET.
\subsection{Preliminaries}
For a large system $Q$, its wavefunction can be arbitrarily bi-partitioned into a \emph{fragment} (or \emph{impurity}) and \emph{bath} (or \emph{environment}).  Some examples of a fragment piece include a single or multiple sites of a lattice model or localized atomic orbitals corresponding to a piece of a larger molecular basis set.  The total wavefunction $\ket{ \psi }$ is naturally expressed in the tensor product of basis states of the fragment and the bath, $\{ \ket{a} \otimes \ket{b} \}$, which has a linear dimension of $d_A \times d_B$ where $d_A \colonequals \dim A$ (resp.\ $d_B\colonequals\dim B$) is the dimension of the Hilbert space of the $A$ fragment (resp.\ $B$ bath). Utilizing the Schmidt decomposition, an eigenstate of $Q$ can be written as a sum over the tensor products of the Schmidt basis states\footnote{This Schmidt basis is constructed by the singular value decomposition of the coefficient tensor.}~\cite{jctc.6b00316.Wouters.Chan, peschel2012special, nielsen2010quantum}.
\begin{align}\label{SVD_wf}
\ket{\psi}
=& \sum_{a}^{d_A}\sum_{b}^{d_B}\psi_{a,b}\ket{a} \ket{b} \nonumber \\
=& \sum_{a}^{d_A}\sum_{b}^{d_B} \sum_{\alpha}^{\min(d_A, d_B)}U_{a,\alpha}\lambda_{\alpha}V_{\alpha, b}^{\dagger}\vert a \rangle \vert b \rangle \nonumber \\
=& \sum_{\alpha}^{\min(d_A, d_B)}\lambda_{\alpha}\vert \tilde{a}_{\alpha}\rangle \vert \tilde{b}_{\alpha}\rangle
\end{align}
This reformulation demonstrates that there is a local fragment basis and environment basis of the same size that produces an equivalent representation of $\ket{\psi}$. Without loss of generality, we can assume the Hilbert space of the fragment is smaller than the environment.  In the Schmidt basis, no matter how large the bath, a fragment $A$ can only entangle with $d_A$ bath states.
The Schmidt states can be used to project the Hamiltonian into a combined impurity/bath basis that necessarily has the same ground state as the original Hamiltonian but is significantly smaller in size:
\begin{align}
\hat{H} \rightarrow \sum_{a,a',b,b'} \ket{ab}\bra{ab} \hat{H} \ket{a'b'}\bra{a'b'}.
\end{align}

For a general large-scale quantum system the wavefunction $\ket{\psi}$ can only be computed approximately.  This constraint naturally leads to the fundamental approximation in DMET: the embedded Hamiltonian is approximated by constructing bath states from the Schmidt decomposition of the ground state of an approximate quadratic Hamiltonian for the total system. As a consequence, the embedded  Hamiltonian now contains an interacting fragment embedded in a non-interacting bath.  The exact embedding Hamiltonian is approximated by matching the fragment's $1$-RDM with the low-level mean-field density matrix of the system by varying the embedding potential that appears in the quadratic Hamiltonian~\cite{PhysRevLett.109.186404}.

\subsection{The Low-Level Quadratic Hamiltonian}\label{sec:lowlevel}
For an arbitrary chemical system, the Hamiltonian contains terms that are at most four creation/annihilation operators corresponding to one- and two-particle interactions:
\begin{align}\label{eq:chem_ham}
\hat{H}_{\textrm{chem}} = \sum_{i,j}\;^{1}h_{i,j}\hat{a}_{i}^{\dagger}\hat{a}_{j} + \frac{1}{2}\sum_{ij,kl}\;^{2}V_{ij,kl}\hat{a}_{i}^{\dagger}\hat{a}_{j}^{\dagger}\hat{a}_{l}\hat{a}_{k}.
\end{align}
In order to vary the form of the bath states used when constructing the embedded Hamiltonian, a potential is added to the full chemical Hamiltonian. The augmented total system Hamiltonian is then approximated using the Hartree--Fock method to find an optimal single-particle basis:
\begin{align}\label{eq:augHF}
\hat{H}_{\textrm{chem}} \rightarrow \hat{F}_{\textrm{chem}} + \sum_{r,s} \; v_{r,s}\hat{a}_{r}^{\dagger}\hat{a}_{s}.
\end{align}
\subsection{Bath Oribtals from a Slater Determinant}
An alternative way to construct the fragment and bath embedding states that is similar to the Schmidt decomposition in \eqref{SVD_wf} is a contraction over the bath basis:
\begin{align}
\ket{\psi}
=& \sum_{a,b}\psi_{a,b}\ket{a}\ket{b}\\
=& \sum_{a} \ket{a} \underbrace{\left( \sum_{b} \psi_{a, b} \ket{b}\right)}_{\colonequals \ket{\tilde a}}\\
=& \sum_{a} \ket{a}\ket{\tilde a}.
\end{align}
Orthogonalizing $\{\ket{\tilde a}\}$ into a set of states $\{\ket{a'}\}$ gives:
\begin{equation}
\ket{\psi} = \sum_{a,a'}\psi_{a,a'}\ket{a}\ket{a'}.
\end{equation}
Note that while each $\ket{a'}\in B$, there are no more than $\dim A$ of them. With this form for the embedding basis, one simply needs to find the form of the bath states constructed from the mean-field wavefunction. This is achieved by projecting the occupied states with overlap on the fragment onto the environment orbitals and normalizing.

There are no formal restrictions on the low-level wavefunction representing the total system $Q$, though it is desirable that one can quickly construct an approximate eigenfunction of $Q$. Alternative approximate wavefunctions that have been used are the antisymmeterized geminal power wavefunctions~\cite{jcp.143.024107.Tsuchimochi.VanVoorhis} that introduce some correlation into the bath, products of Bogoliubov quasi-particle states~\cite{PhysRevB.93.035126}, and a Slater determinant~\cite{PhysRevLett.109.186404}. The following is a derivation of the embedding basis from a Slater determinant wavefunction.  This low-level wavefunction for the full system $\ket{\phi_{0}}$ can be expressed as the product of fermionic operators in the occupied subspace acting on the true vacuum:
\begin{align}
\ket{ \phi_{0} } = \Big(\prod_{p} \hat{a}_{p}^{\dagger}\Big)\ket{ \textrm{vac} }.
\end{align}
The single particle orbitals $\hat{a}_{p}^{\dagger}$ can be constructed from linear combinations of site orbitals, $\hat{c}^{\dagger}$, in the lattice:
\begin{align}
\hat{a}_{p}^{\dagger} = \sum_{\mu}D_{\mu, p}\hat{c}_{\mu}^{\dagger}.
\end{align}
The matrix $D$ corresponds with the Hartree--Fock transform from local lattice spin-orbitals to the basis that minimizes the Hartree--Fock Hamiltonian of the system. A rotation to the fragment and bath basis can be constructed in a similar manner to the canonical orthogonalization of atomic orbitals that appears in the Hartree--Fock procedure applied to molecules~\cite{szabo1989modern, helgaker2014molecular}.

The overlaps of the occupied orbitals projected onto the fragment space are calculated using the fragment projection operator $\hat{P}_{F} \colonequals \sum_{\mu_{F}}\ket{ \mu } \bra{ \mu }$ acting on the occupied set of orbitals indexed by $p$ and $q$:
\begin{align}
S
&= \braket{ \hat{P}_{F} \phi_{p} \mid \hat{P}_{F} \phi_{q} } \nonumber \\
&= \bra{ \phi_{p} } \hat{P}_{F}^{\dagger}\hat{P}_{F} \ket{ \phi_{q} } \nonumber \\
&= \bra{ \phi_{p} } \hat{P}_{F} \ket{ \phi_{q} }.
\end{align}
The eigenvectors of this overlap matrix correspond to an orbital rotation that forms the fragment and bath basis. Diagonalizing $S$ with a unitary $V$ gives
\begin{equation}
\Delta \colonequals V^{\dagger}SV = \diag(\Delta_1, \ldots, \Delta_{d_A}).
\end{equation}
Naturally, the eigenvalues of the overlap matrix are between $0$ and $1$. Zero eigenvalues correspond to occupied states that have zero overlap on the fragment. The bath states can be constructed by projecting the occupied orbitals with overlap on the fragment into the bath:
\begin{equation}\label{bathstates}
\ket{ b_{i} } = \sum_{p} \frac{V_{p,i}^{*}}{\sqrt{1 - \Delta_{i}}} \hat{P}_{B}\ket{ \phi_{p} }.
\end{equation}
Finally, the rotation $C$ to the Schmidt basis is defined by the direct sum of the bath transformation and an identity matrix of dimension equal to the number of orbitals on the fragment, $n_{F}$:
\begin{equation}\label{eq:overlap}
C \colonequals I_{n_F} \oplus B = \diag(I_{n_F}, B).
\end{equation}
The core states with zero overlap on the fragment are eliminated from the embedded Hamiltonian by including their interaction at the mean-field level similar to generating an active space interacting with a frozen set of core states~\cite{Shavitt1.430426}.
\subsection{Embedded Hamiltonian}
The embedded Hamiltonian is constructed by projecting into the fragment and bath states
\begin{align}
\hat{H}_{\text{emb}} \colonequals P\hat{H}P
\end{align}
which, for a Slater determinant wavefunction, corresponds to a single- and double-particle integral transform~\cite{jctc.301044e.Kinzia.Chan}:
\begin{align}\label{eq:embH}
\hat{H}_{\text{emb}} =& \sum_{p, q} \;^{1}\tilde{h}_{p, q}\hat{a}_{p}^{\dagger}\hat{a}_{q} + \frac{1}{2}\sum_{p, q; r, s}\;^{2}\tilde{V}_{p,q; r,s}\hat{a}_{p}^{\dagger}\hat{a}_{q}^{\dagger}\hat{a}_{s}\hat{a}_{r}\\
\shortintertext{where}
&^{1}\tilde{h}_{p,q} \colonequals C^{\dagger} \;^{1}h_{a,b} C  + f_{pq}^{\textrm{core}} \\
&^{2}\tilde{V}_{p,q;r,s} \colonequals \left( C \otimes C \right)^{\dagger} \; ^{2}V_{a,b; c,d} \left(C \otimes C\right).
\end{align}
In the above transformation $f^{\textrm{core}}$ is the interaction with the non-entangled non-overlapping core states, $C$ is a matrix from \eqref{eq:overlap} whose columns are the transformation vectors to the embedding basis, and ${}^{1}h$ and ${}^{2}V$ are the one- and two-particle integral tensors defined in the lattice basis.

\subsection{Determining the Optimal Embedding Potential}
The key step in the DMET procedure requires improving the bath states so they approximate the bath states of the true embedded Hamiltonian.  Schmidt bath state tuning is achieved by varying the potential that was added to the system Hamiltonian in Sec.~\ref{sec:lowlevel}.  Variation in this potential is the connection between the embedded Hamiltonian and the quadratic Hamiltonian representing the entire system.  The total DMET procedure can be summarized as follows:
\begin{enumerate}
    \item Find the ground state Slater determinant wavefunction $\vert \phi \rangle$ for the entire system $Q$ with the Hartree-Fock Hamiltonian augmented with the embedding potential.
    \item For all fragments, use $\ket{\phi}$ to construct the embedding basis and project the system Hamiltonian into each embedded Hamiltonian.  Solve for the $1$-RDM and $2$-RDM with a high-level method.
    \item Adjust the embedding potential in the mean-field Hamiltonian such that the $1$-RDM of the impurity calculated with the high-level wavefunction matches the $1$-RDM calculated from the quadratic Hamiltonian.
    \item Repeat until particle conservation and the embedding potential do not change between iterations.
\end{enumerate}
Step 3 is carried out by minimizing the squared norm between the fragment piece of the mean-field $1$-RDM and the fragment $1$-RDM computed from the ground state wavefunction for the embedded Hamiltonian:
\begin{align}
\mathrm{CF}_{\textrm{frag}}(u) = \sum_{x}\sum_{r,s \in \mathrm{frag}} \left(D_{r,s}^{x} - D_{r,s}^{\textrm{mf}}(u)\right)^{2}.
\end{align}
In \cite{jctc.6b00316.Wouters.Chan} this cost function is known as the \textit{fragment-only} density matrix fitting.  Alternatives such as full matching of the $1$-RDMs in the embedding basis or matching the electron densities on the fragment have also been studied~\cite{PhysRevB.89.035140, jcp.141.054113.Ireneusz.Scuseria}.

The exponentially scaling FCI has prompted studying alternative methods for finding the correlated ground state $1$-RDM~\cite{jctc.6b00316.Wouters.Chan, zheng2016cluster}.  Though this may not lead to an optimal approximation to the embedded Hamiltonian fragment and bath states it does allow for the study of significantly larger fragments--important for assessing longer range correlation functions and more accurate local expectation values~\cite{PhysRevB.93.035126}.  The ability to study larger fragments at higher fidelity has many implications for accurate simulations of materials and chemistry.  For example, if one were studying a reaction center in a metal-organic framework a fragment would likely need to contain the metal site and ligand orbitals for an accurate description of the energy and properties.  The VQE enters in step $2$ when solving for the $1$-RDM and $2$-RDM of the embedded Hamiltonian.  DMET's need for the $1$-RDM of the embedded Hamiltonian makes VQE an ideal quantum algorithm for this use case.

\section{The Unitary Cluster Ansatz and Trotterization in VQE}
The VQE is a functional minimization scheme that leverages fast construction of the wavefunction to find expectation values of operators~\cite{PhysRevX.6.031007, mcclean2016theory}.  The energy $E$ of the system is minimized by varying over parameters for the wavefunction ansatz \eqref{QVE_variational}:
\begin{align}\label{QVE_variational}
E = \min_{\theta} \bra{ \psi(\theta) } \hat{H} \ket{ \psi(\theta) }.
\end{align}
The expectation value is determined by summing the expectation of each term in the Hamiltonian:
\begin{equation}
\langle\hat{H}\rangle = \sum_{k}\langle \hat{H}_{k} \rangle.
\end{equation}
For a general quantum chemical Hamiltonian, $k$ scales quadratically with respect to basis set size.  In order to evaluate the Hamiltonian expectation value, the wavefunction $\ket{\psi(\theta)}$ is prepared many times with a subsequent measurement of the $\hat{H}_{k}$ operator.  In this work the unitary coupled cluster (UCC) state is used as the function for the energy functional though others could be considered such as the variational adiabtic ansatz.  

UCC is a many-body expansion wavefunction ansatz parameterized by the cluster coefficients associated with each generator.  We use the UCC-singles-doubles ansatz corresponding to anti-Hermitian generators performing single- and double-particle excitations and de-excitations:
\begin{align}
\ket{ \psi(\theta) } &= \exp\left(\sum_{k=1}^{2}T_k(\theta)\right) \ket{ \psi_{\textrm{ref}} }\label{eq:wf}\\
\shortintertext{where}
T_{1}(\theta) &\colonequals \sum_{a, i}\theta_{a}^{i}\left(\hat{a}_{i}^{\dagger}\hat{a}_{a} - \hat{a}_{a}^{\dagger}\hat{a}_{i}\right)
\label{cluster_ops_tau1}\\
T_{2}(\theta) &\colonequals \sum_{a,b; i,j}\theta_{a,b}^{i,j}\left(\hat{a}_{i}^{\dagger}\hat{a}_{a}\hat{a}_{j}^{\dagger}\hat{a}_{b} - \hat{a}_{b}^{\dagger}\hat{a}_{j}\hat{a}_{a}^{\dagger}\hat{a}_{i}\right).
\label{cluster_ops_tau2}
\end{align}
In quantum computing, the antisymmetric fermionic creation/annihilation operators are represented by distinguishable qubits. We utilize the Jordan--Wigner transformation (JW) for mapping fermionic creation/annihilation operators to Pauli spin-operators that preserve the anti-commutation relations and parity of the second quantized operators \eqref{JW}~\cite{whitfield2011simulation}.  It should be noted that the representation of second quantized operators in Pauli matrices by the JW transform scales linearly in the number of Pauli terms\footnote{There are alternatives such as the Bravyi--Kitaev transformation that have logarithmic scaling when representing second quantized operators~\cite{Tranter.BKTransform.QUA24969, seeley2012bravyi}.} as the basis index is increased. The maps are defined as
\begin{equation}\label{JW}
\hat{a}_{q}^{\dagger} \mapsto \Big(\bigotimes_{i < p} \sigma_{i}^{z} \Big)  \sigma_{p}^{+}
\quad\text{and}\quad
\hat{a}_{q} \mapsto \Big(\bigotimes_{i < p} \sigma_{i}^{z} \Big) \sigma_{p}^{-}
\end{equation}
where
\begin{equation}\label{JW_sigma}
\sigma^{\pm} \colonequals \frac{1}{2}\left(\sigma^{x} \mp i\sigma^{y}\right).
\end{equation}
For each shot of the VQE algorithm the wavefunction of the system---according to a UCC ansatz---is constructed by first preparing the Hartree--Fock reference, $O(n)$ in the number of gates required to prepare the state where $n$ is the number of orbitals, then evolving the initial wavefunction according to \eqref{eq:wf}, using $O(n^{4})$ in the number of gates for the exponentiation of the unitary generators.

The $1$- and $2$-RDM are measured after an optimal set of $\theta$ parameters are determined by measuring $n(n + 1)/2$ expectation values corresponding to the $2$-RDM:
\begin{align}
^{2}D_{kl}^{ij} = \bra{\psi} \hat{a}_{i}^{\dagger} \hat{a}_{j}^{\dagger}\hat{a}_{l}\hat{a}_{k} \ket{\psi}.
\end{align}
The $1$-RDM is obtained by contraction from the $2$-RDM:
\begin{align}
^{1}D_{i}^{j} = \frac{1}{n - 1}\sum_{a}\;^{2}D_{i,a}^{j, a}.
\end{align}
The fragment $1$-RDM of the total system is matched to the fragment piece of the $1$-RDM obtained from the VQE algorithm by varying the embedding potential in \eqref{eq:chem_ham}.

\subsection{Commutation Relations for Generators in Unitary Coupled-Cluster}
Unlike the classical cluster operators that commute at all orders~\cite{crawford2000introduction} the unitary coupled-cluster operators \eqref{cluster_ops_tau1}--\eqref{cluster_ops_tau2}, henceforth denoted as $\tau$, do not commute with each other at all orders.  Therefore approximating the exponential by a particular order of the Suzuki--Trotter decomposition~\cite{suzuki1994convergence, trotter1959product} may be important for constructing accurate distributions. The anti-Hermitian generators do not commute because the de-excitation operators $-\hat{a}_{a}^{\dagger}\hat{a}_{i}$ do not commute with the excitation operators.  This can be seen for the first order $\tau$ operators.  Here we use $a, b$ to denote the particle space and $i, j$ to denote the hole space.
\begin{align}
\left[ \tau_{i}^{a}, \tau_{j}^{b} \right]
=& \left[ \big( \hat{a}_{a}^{\dagger}\hat{a}_{i} - \hat{a}_{i}^{\dagger}\hat{a}_{a} \big) , \big( \hat{a}_{b}^{\dagger}\hat{a}_{j} - \hat{a}_{j}^{\dagger}\hat{a}_{b} \big) \right] \\
=& \left[\hat{a}_{a}^{\dagger}\hat{a}_{i}, \hat{a}_{b}^{\dagger}\hat{a}_{j}\right] + 
\left[\hat{a}_{j}^{\dagger}\hat{a}_{b}, \hat{a}_{a}^{\dagger}\hat{a}_{i} \right]\nonumber\\  
&\phantom{=} + \left[\hat{a}_{b}^{\dagger}\hat{a}_{j}, \hat{a}_{i}^{\dagger}\hat{a}_{a}\right] + 
\left[  \hat{a}_{i}^{\dagger}\hat{a}_{a},  \hat{a}_{j}^{\dagger}\hat{a}_{b}\right] 
\end{align}
The first and last commutators vanish while the middle two terms survive when there is an overlapping index in $\tau_{i}^{a}$ and $\tau_{j}^{b}$:
\begin{align}
\left[\hat{a}_{j}^{\dagger}\hat{a}_{b}, \hat{a}_{a}^{\dagger}\hat{a}_{i} \right]  &=  \delta_{b,a}\hat{a}_{j}^{\dagger}\hat{a}_{i} - \delta_{j,i}\hat{a}_{a}^{\dagger}\hat{a}_{b}\\
\left[\hat{a}_{b}^{\dagger}\hat{a}_{j}, \hat{a}_{i}^{\dagger}\hat{a}_{a}\right] &= \delta_{i, j}\hat{a}_{b}^{\dagger}\hat{a}_{a} - \delta_{b,a}\hat{a}_{i}^{\dagger}\hat{a}_{j}.
\end{align}
As a result, Trotter error must be considered when generating circuits for UCC state preparation.  In Sec.~\ref{sec:results} we discuss the performance of UCCSD with Trotter orders $1$ and $2$ with Trotter steps beyond $1$.\footnote{The term \emph{Trotter order} refers to the order of the series approximation to the exponential of two non-commuting operators.  The term \emph{Trotter steps} refers to the number of slices that each order is broken into in order to minimize the the correction term.}

\section{Results and Discussion}\label{sec:results}
\begin{table*}[ht]
    \newcolumntype{.}{D{.}{.}{-1} }
    \centering
    \caption{Energy-per-site of the half-filled Hubbard model for a four-site ring with anti-periodic boundary conditions evaluated with DMET using one- and two-site fragments. The exact solution is computed with an exact diagonalization of the Hamiltonian. Both Trotter order and Trotter steps were equal to one for all UCCSD calculations\label{resultTab1}}
    \begin{tabular}{l.....}
    \hline
    \hline
    \multicolumn{1}{c}{$U$} & 2 & 4 & 6 & 8 & 10\\
    \hline
    Exact      &   -0.9809782 & -0.68014156 & -0.49157349 & -0.37607898 & -0.30214434 \\
    UCCSD      &   -0.9808687 & -0.67928156 & -0.48543800 & -0.33051713 & -0.02603051 \\
    DMET(1)-ED &   -0.9951259 & -0.71791138 & -0.54055767 & -0.42535625 & -0.34751768 \\
    DMET(2)-ED &   -0.9808783 & -0.68014156 & -0.49157349 & -0.37607898 & -0.30214434 \\
    DMET(1)-UCCSD &-0.9951259 & -0.71791135 & -0.54055772 & -0.42535622 & -0.34751654 \\
    \hline
    \hline
    \end{tabular}
\end{table*}
We considered the one-dimensional Hubbard model \eqref{eq:hubbard_hamiltonian} with repulsive-$U$ interactions and anti-periodic boundary conditions as a representative test system for DMET.
\begin{align}\label{eq:hubbard_hamiltonian}
\hat{H} =& -t\sum_{\langle i, j\rangle}\left(\hat{a}_{i, \sigma}^{\dagger}\hat{a}_{j, \sigma} + \textrm{h.c.} \right)\\
&\phantom{=} + U\sum_{i}\hat{a}_{i, \alpha}^{\dagger}\hat{a}_{i, \beta}^{\dagger}\hat{a}_{i, \beta}\hat{a}_{i, \alpha}\nonumber
\end{align}
The DMET self-consistency loops were run using the QC-DMET~\cite{QC-DMET} code with an interface to PySCF~\cite{PySCF} for exact diagonalization of the embedded Hamiltonian.  The VQE algorithm for solving the embedded Hamiltonian used Quil and Rigetti Computing's pyQuil package to construct the UCCSD ansatz and evolved the wavefunction on a noiseless quantum virtual machine~\cite{smith2016practical}. We describe the programming environment in more detail in Appendix~\ref{sec:progenv}.  All UCCSD-VQE runs started with a second-order M{\o}ller--Plesset perturbation theory guess for the cluster amplitudes and required about 20 iterations of BFGS~\cite{nocedal2006numerical} with the gradient numerically approximated. For single-site DMET calculations the low-level and high-level fragment density matrices were equivalent and thus the DMET loop only involved setting the chemical potential such that the number of electrons in each fragment summed to the total system's electron count.  For two-site models the fragment $1$-RDM was generally not equal to the fragment Slater determinant $1$-RDM and thus required a numerical search for the optimal embedding potential.

\begin{figure}
    \centering
    \includegraphics[width=8.5cm]{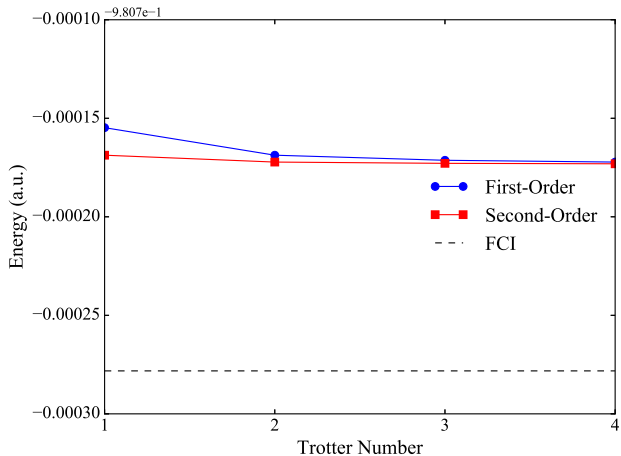}
    \caption{Energy-per-site of the four-site Hubbard model with $U/|t| = 2$ with anti-periodic boundary conditions calculated with Trotter orders 1 and 2 with Trotter steps 1, 2, 3, and 4.  The error between Trotter steps of Suzuki--Trotter orders is generally hundredths of a milli-Hartree indicating that the significant gap between FCI and UCCSD is based on the wavefunction ansatz.}
    \label{fig:TrotterResponse}
\end{figure}

The non-commuting nature of the cluster operators do not drastically effect the accuracy of the UCCSD ansatz.  Figure~\ref{fig:TrotterResponse} plots the convergence of the energy-per-site as the Trotter order and Trotter steps are increased for $U/|t| = 2$.  Trotter order and Trotter steps indicate the structure of the wave function ansatz for UCCSD.  For example, the first order Trotter with $N$-Trotter steps produces an ansatz of form:
\begin{align}
\vert \psi \rangle = \left(e^{\hat{\tau}_{s}/N}e^{\hat{\tau}_{d}/N}\right)^{N}
\end{align}
where $\tau_{s}$ and $\tau_{d}$ correspond to the single and double anit-Hermitian generators with their coefficients which are described in Eq.~\eqref{cluster_ops_tau1} and Eq.~\eqref{cluster_ops_tau2}.
The persistence of the energy gap as Trotter order and Trotter steps are increased demonstrates that the single reference nature and the truncated set of generators are the main sources of error. 

The DMET methodology introduces three sources of error:
\begin{inlinelist}
\item the fragment size and thus the number of non-interacting bath states that can entangle with the fragment,
\item the embedded Hamiltonian solver, and
\item the pieces of the embedded Hamiltonian $1$-RDM that are chosen to match for updating the embedding potential.
\end{inlinelist}
The effects of (i) and (ii) are determined by studying small lattices and comparing the ground state energy against the FCI solution and the full UCCSD solution.  Table~\ref{resultTab1} compares the ground state energy-per-site of a four-site ring at various interaction strengths between the FCI solution, the UCCSD solution, and various fragment sizes for DMET.  The DMET solutions are labeled by DMET($n$), where $n$ is the number of sites considered in the fragment, followed by an acronym for the type of embedded solver--i.e. ED corresponds to FCI.  The accuracy of the UCCSD solution to the four-site-ring lattice generally decreases as $U$ is increased demonstrating the expected result that singles and doubles generators cannot parametrize the full unitary group.  A four-site lattice requires eight-spin-orbitals and thus eight qubits.  The UCCSD ansatz contains 14 cluster amplitudes when restricting the wavefunction to the singlet subspace~\cite{Bulik.jctc.5b00422, Scuseria.jcp.1.455269}.   Under a greedy parallelization of commuting instructions for breaking the unitary evolution into time steps, a single state evolution at first order Trotter involves $1422$ time steps where multiple qubit operations are performed at each step.  Each time step has an average of 0.8692 one-qubit gates and 0.8101 two-qubit gates. The coupled-cluster reference is usually chosen in a basis where the one-particle piece of the Hamiltonian is diagonal.  For the Hubbard model this involves a Fourier transform on the one-particle and two-particle integral tensors resulting in $m$ one-body terms and $m^{3}$ two-body terms where $m$ is the number of lattice sites.  The two-body terms are proportional to $U/N$.

A single fragment site results in a trivial two-orbital embedded Hamiltonian.  For this system, UCCSD + VQE is equivalent to FCI, CISD, and CCSD solvers.  Considering a two-site DMET at $U/|t| = 2$ and $U/|t| = 8$ produced energies-per-site of $-0.9809165$ and $-0.35446392$, respectively.  Though not exactly equal to the DMET(2)-ED result, the error is on the order of the difference between the UCCSD solution for the full lattice and the FCI solution for the full lattice.

\begin{figure}
    \centering
    \includegraphics[width=8.5cm]{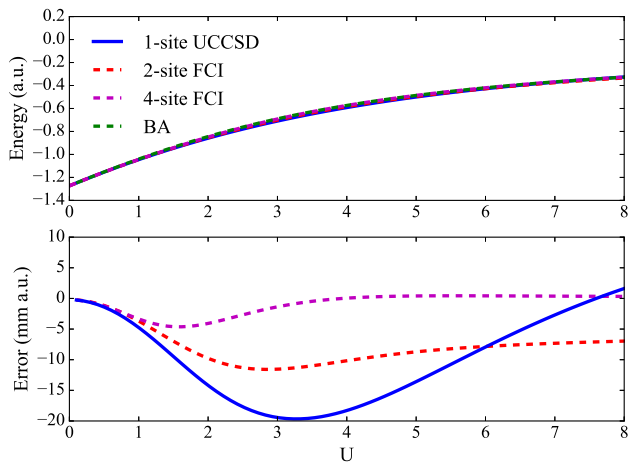}
    \caption{\textit{top}: Energy-per-site in the 100-site Hubbard model with a one-, two-, and four-impurity sites in the DMET scheme along with the Bethe solution described in \cite{PhysRevB.6.930}.  FCI is the exact diagonalization solution to the single impurity site problem. \textit{bottom:} The energy difference between the Bethe ansatz solution and the DMET solutions.  As expected, the error decreases significantly as the number of sites increases.  The number of qubits required for each embedded Hamiltonian eigenvalue problem is four times the number of sites included in the fragment. \label{fig:hubbard_energy_half}}
\end{figure}
In Figure~\ref{fig:hubbard_energy_half}, we plot the energy-per-site for a $100$-site lattice calculated with DMET and the Bethe ansatz~\cite{lieb2003one, PhysRevB.6.930}.  The top subfigure demonstrates that DMET(1)-UCCSD method converges to the correct solution at high-$U$.  The bottom subfigure describes the error for DMET(1)-ED, DMET(1)-UCCSD, DMET(2)-ED, and DMET(4)-ED relative to the Bethe ansatz.  As expected the quality of the solution improves as fragment size is increased.  An $m$ site fragment requires $4m$ qubits and thus we are restricted to small fragment sizes when performing a classical simulation. 

\section{Conclusion}
The hybrid classical/quantum computation model naturally fits with current state-of-the-art methods for studying the electronic structure of molecules and materials.   Many of the current methods involve selecting an active space or embedded piece of the Hamiltonian to treat at higher fidelity.  These techniques are especially important when describing the electronic structure of correlated materials and molecules such as metal-oxides, heterogeneous catalysis, multi-electron redox reactions, and catalysts.  In contrast to exact diagonalization solvers commonly used in impurity or active space problems, quantum computation offers a technique to treat these Hamiltonians with high accuracy in polynomial time.   

The first steps towards integrating a quantum computational resource with embedding schemes was outlined in \cite{bauer2015hybrid}.  That work integrated quantum phase-estimation with DMFT as the impurity solver.  Their results indicate that a modestly sized quantum computational resource of about 100 qubits can be used to simulate a relevant physical problem.  Though their work describes DMFT as the impurity methodology they do discuss the use of alternative embedding schemes along with alternative algorithms to use on the quantum computation side.

In this work we have elaborated on one alternative in particular: DMET integrated with the VQE.  DMET works by mapping the problem of finding the $N$-particle ground state of a large system to many smaller interacting impurity problems.  Therefore, DMET is applicable to very large scale problems be it either molecules or materials.   Unlike DMFT that requires the frequency dependent two-particle Green's function, DMET requires the $1$-RDM of the embedded problem with high accuracy.  This requirement makes DMET and the VQE a natural pair.  The $1$-RDM and $2$-RDM are computed at each iteration of the VQE optimization.  

The VQE algorithm is a general functional minimization routine that allows for the use of any quantum state preparation method.  In this work we utilized the UCC ansatz and demonstrated that despite the non-commuting nature of the generators, the error for the ansatz comes from the truncation of the series and not Trotter error.  Therefore, at least in the case of simulating local interaction models, low-order Trotter approximation and Trotter number are sufficient for accurate state preparation.  

Current classical techniques for describing quantum chemistry or correlated materials are naturally accelerated with a hardware quantum processing unit (QPU).  Many of the classical methods that require a $1$-RDM or $2$-RDM of an active space can be integrated with the VQE algorithm or other hybrid classical/quantum techniques. Further integration can lead to methodologies to study large quantum systems with higher resolution than what is currently possible with state of the art classical methods.

\section{Acknowledgments}
The author acknowledges useful discussions of VQE and DMET with Jarrod McClean and members of the Rigetti Computing software team: Will Zeng and Robert Smith.  A special thanks to Robert Smith for writing the Appendix on quantum programming and detailed editing of the draft.

\section{Appendix}

\subsection{Details of the Quantum Programming Environment}\label{sec:progenv}
In this work, execution of quantum programs was done using the quantum abstract machine and Quil, its quantum instruction language \cite{smith2016practical}. In this section, we outline elements of this quantum programming environment, including examples of essential techniques used to compute the $1$- and $2$-RDMs.

The VQE is, at its core, a classical optimizer around a quantum execution unit. In particular, a classical optimization loop produces the next $\theta$ parameters of \eqref{QVE_variational} for minimization. Mathematically, $\theta$ parameterizes a wavefunction ansatz, but computationally, they are regarded as parameters to generate a new quantum program. These programs are run on a quantum execution unit modeled as a restricted quantum abstract machine\footnote{The machine $(\ket{\Psi}, C, G, G', P, \kappa)$ is one with quantum state $\ket{\Psi}$, classical state $C$, static gates $G$, parametric gates $G'$, program $P$, and program counter $\kappa$.} $\mathfrak{M} = (\ket{\Psi}, C, G, G', P, \kappa)$ with
\begin{align}
G &= \left\{\sigma^x, \sigma^y, \sigma^z, \mathsf{H},
\tfrac{1}{\sqrt{2}}\left(\begin{smallmatrix}
1 & \pm i\\\pm i & 1
\end{smallmatrix}\right),
\mathsf{CNOT}
\right\}\\
G' &= \{\mathsf{R}_z\},
\end{align}
where the gates are understood to be the collection of operators acting on each qubit combination of the quantum state $\ket{\Psi}$. The classical memory $C$ is as large as $\log_2\dim\ket{\Psi}$---the number of qubits---to hold the $z$-basis measurements for each loop of the VQE, which are collected to approximate $\langle\hat H\rangle$. Alternatively, by using a quantum virtual machine, the amplitudes of $\ket{\Psi}$ can be observed in the computational basis and $\langle\hat H\rangle$ can be computed directly.

The object of interest to simulate is usually specified in terms of sums of fermionic operations, such as the Hamiltonian in \eqref{eq:embH}. After a fermionic transform such as~\eqref{JW}, said object is represented as a sum of products of Pauli spin operators. We represent these Pauli sums algebraically as first-class objects in a canonical form. We wish to compute a program which acts identically to the exponentiation of this Pauli representation---as in \eqref{eq:wf}---on the quantum abstract machine's $\ket{\Psi}$. Exponentiating a Pauli sum requires two steps: factorization via Trotterization, and exponentiation of Pauli terms. Consider the Pauli terms
\begin{equation}
A = 2\sigma^x_0\sigma^x_1\quad\text{and}\quad
B = -\tfrac{1}{2}\sigma^x_0\sigma^z_2.
\end{equation}
These are represented as objects in Python using the pyQuil library:
\begin{verbatim}
> A = 2.0*PauliTerm('X', 0)*PauliTerm('X', 1)
> B = -0.5*PauliTerm('X', 0)*PauliTerm('Z', 2)
> print "A =", A, "\nB =", B
A = 2.0*X0*X1
B = -0.5*X0*Z2
\end{verbatim}
Exponentiation of a Pauli term has three parts: a change to the $\ket{\pm}$-basis, an entanglement of the qubits on which the operator non-trivially acts, and a rotation about $z$. For example, we can see this by looking at a Quil program which allows one to compute the action of $e^A$.
\begin{verbatim}
> print exponentiate(A)
H 0
H 1
CNOT 0 1
RZ(4.0) 1
CNOT 0 1
H 0
H 1
\end{verbatim}
This Quil program itself is also a first-class object. Indeed, the result of a Suzuki--Trotter approximation of, say, $e^{A+B}$ must combine programs which compute $e^A$ and $e^B$. We can see this by looking at the Quil program which computes $e^{A+B}$ in a first-order decomposition:
\begin{verbatim}
> print trotterize(A, B)
H 0
H 1
CNOT 0 1
RZ(4.0) 1
CNOT 0 1
H 0
H 1
H 0
CNOT 0 2
RZ(-1.0) 2
CNOT 0 2
H 0
\end{verbatim}

Quil programs are written as a serial list of instructions, but in fact the semantics are not changed if the order between successive commuting instructions is changed. As such, we can think of such instructions occurring in a single \emph{time slice}\footnote{If the $n$\textsuperscript{th} commuting instruction executes in $t_n$ time, then the time slice itself executes in $\max_n t_n$ time and is the worst case for arbitrary $t_n$ with maximal parallelization. If all $t_n$ are known for each time slice, then it is possible to do \emph{aggressive parallelization} by overlapping commuting instructions from subsequent time slices, despite the time slices as a whole (i.e., its action on the total Hilbert space) being non-commuting.}. A straight-line Quil program itself is \emph{parallelized} if it is transformed according to a maximal semantics-preserving parallelization. Consider the following 16-instruction sample from a state evolution according to the UCCSD ansatz with first-order Trotter and one time slice for the four-site Hubbard model:
\begin{verbatim}
X 2
X 3
H 3
RX(1.5707963267948966) 5
CNOT 3 4
CNOT 4 5
RZ(0.00001) 5
CNOT 4 5
CNOT 3 4
H 3
RX(-1.5707963267948966) 5
RX(1.5707963267948966) 1
H 5
CNOT 1 2
CNOT 2 3
CNOT 3 4
\end{verbatim}
Instruction sequences like \verb|X 2| and \verb|X 3| commute and may execute in the same time slice. However, \verb|X 3| and \verb|H 3| do not commute and must execute in different time slices. The parallelized Quil program is as follows:
\begin{verbatim}
Time Slice #1:  X 2
                X 3
                RX(1.5707963267948966) 5
                RX(1.5707963267948966) 1
Time Slice #2:  H 3
                CNOT 1 2
Time Slice #3:  CNOT 3 4
Time Slice #4:  CNOT 4 5
Time Slice #5:  RZ(0.00001) 5
Time Slice #6:  CNOT 4 5
Time Slice #7:  CNOT 3 4
                RX(-1.5707963267948966) 5
Time Slice #8:  H 3
                H 5
Time Slice #9:  CNOT 2 3
Time Slice #10: CNOT 3 4
\end{verbatim}
The 16-instruction straight-line program has been reduced to a 10-time-slice program averaging $0.9$ one-qubit gates and $0.7$ two-qubit gates per time slice. All instructions being equal, this gives a 38\% improvement in timing.

When a final Quil program is prepared through the various means described, it is dispatched to a representation of a QAM: either a hardware QPU or a software quantum virtual machine. In this work, a remotely deployed quantum virtual machine was used for all Quil execution. 
\bibliography{biblo}
\end{document}